\documentclass{llncs} 

\usepackage[T1]{fontenc}
\usepackage{epstopdf}
\usepackage{float,url}
\usepackage[caption=false,font=footnotesize]{subfig}
\usepackage{amsmath,amssymb} 
\usepackage{amsfonts,graphics,graphicx}
\usepackage{algorithm}
\usepackage{algorithmicx} 
\usepackage{algpseudocode}
\usepackage{textcomp}
\usepackage{xcolor}
\usepackage{multirow}

\newcommand{\ignore}[1]{}
\newtheorem{insight}{Insight}
\newcommand{\A}{\mathcal{A}}
\newcommand{\N}{\mathbb{N}}

\begin{document}




\title{AutoCRAT: Automatic Cumulative Reconstruction of Alert Trees}


\ignore{
\author{Eric Ficke}
\affiliation{%
\institution{The University of Texas at San Antonio}
}

\author{Raymond M. Bateman} 
\affiliation{\institution{U.S. Army Research Laboratory South - Cyber}}

\author{Shouhuai Xu} 
\affiliation{%
\institution{University of Colorado Colorado Springs}
}
}



\author{Eric Ficke\inst{1}\orcidID{0000-0002-3762-6475} \and 
Raymond M. Bateman\inst{2}\orcidID{0000-0002-2949-5145} \and 
Shouhuai Xu\inst{3}\orcidID{0000-0001-8034-0942}}

\institute{The University of Texas at San Antonio, San Antonio, TX, USA \and
U.S. Army Research Laboratory South - Cyber, San Antonio, TX, USA \and
University of Colorado Colorado Springs, Colorado Springs, CO, USA}


\maketitle 

\begin{abstract}
When a network is attacked, cyber defenders need to precisely identify which systems (i.e., computers or devices) were compromised and what damage may have been inflicted. This process is sometimes referred to as {\em cyber triage} and is an important part of the incident response procedure. Cyber triage is challenging because the impacts of a network breach can be far-reaching with unpredictable consequences. This highlights the importance of {\em automating} this process. In this paper we propose AutoCRAT, a system for quantifying the breadth and severity of threats posed by a network exposure, and for prioritizing cyber triage activities during incident response. Specifically, AutoCRAT automatically reconstructs what we call {\em alert trees}, which track network security events emanating from, or leading to, a particular computer on the network. We validate the usefulness of AutoCRAT using a real-world dataset. Experimental results show that our prototype system can reconstruct alert trees efficiently and can facilitate data visualization in both incident response and threat intelligence analysis.
\end{abstract}



\section{Introduction}
In cyber incident response, the defender needs to precisely identify what happened to the network in question, including: how did the attacker propagate through the network, what was the attacker's intent, and where and how much damage did the attacker inflict? Since attackers may target a large portion of a network, the defender must quickly and effectively determine the scope of their impact. Specifically, the defender must isolate the routes that the attacker may have used to enter and propagate through the network. These are referred to as {\em alert paths}, and may be aggregated into so-called {\em alert trees}. 

Isolating alert paths turns out to be a difficult task for two reasons. First, for any amount of incoming alerts, the number of paths that need to be examined grows quadratically. This is the problem of {\em efficiency}.
Second, without examining all possible alert paths, it is possible that the defender will overlook the actual attack path. This is the problem of {\em coverage}, which is closely related to false negatives in intrusion detection systems.
These problems have serious implications on incident response, 
as the average time to contain threats can be as high as 85 days \cite{ibm2021breachreport}. The same report suggests that security automation can reduce this by 26\% (to 63 days). In order to help defenders effectively and efficiently respond to cyber incidents, the research community needs to investigate principled solutions to tackling this problem. 
This motivates the present study, which aims to facilitate {\em cyber triage} by automatically identifying the potential scope of an attack.


\ignore{
Concepts - Alert tree (derived from attack tree/path, as empirical concept rather than predictive/probabilistic
Methods - maintain set of chronological paths, merge on demand to form trees
Architecture - as shown
Algorithms - as described
Heuristic - Threat Score (ETS/PTS derived from ITS and CTS)
Case Study - CSE-CIC-IDS2018
}

\noindent{\bf Our Contributions}.
In this paper, we make three contributions.
First, we formalize a suite of concepts to automate cyber triage, including: {\em alert graph}, which presents a Graph-Theoretic representation of alerts triggered on a network; {\em alert path}, which indicates a sequence of potentially-related attack steps across multiple computers; {\em alert tree}, which represents a set of attack paths (as identified by alerts) with respect to a computer of interest.
These concepts allow us to represent alerts in a structural fashion to facilitate various analysis and reasoning cyber triage tasks of interest. 

Second, we present a novel system called AutoCRAT to reconstruct alert trees from the output of a Network Intrusion Detection System (NIDS). AutoCRAT's key features can be characterized as follows.
(i) 
It can continuously process streams of alerts reported by security devices. This is important for real-world usage where security devices constantly produce alerts.
(ii) 
It can quickly reconstruct alert trees on demand. Our asymptotic analysis shows that in the worst-case scenario, graph maintenance scales cubically with the number of alerts, while tree reconstruction scales quadratically or log-linearly, depending on the type of tree. 
(iii) It can model {\em multi-step attacks}, which include lateral movements secondary to a compromise. This is important because attacker objectives often cannot be accomplished after a single compromise. 
(iv) It can model {\em muti-target attacks}, in which a single attacker produces multiple attack paths with different targets. This is important because it does not require the assumption that an attacker has a single target. 
(v) It can model {\em multi-source attacks}, in which a single victim is targeted by different attackers or by an attacker with multiple access points. This is important because it does not require the assumption that only a single attacker is active on the network. 

Third, we demonstrate the usefulness of alert trees and the alert tree reconstruction method by conducting a case study using a dataset which is collected from a realistic cyber attack testbed, namely CSE-CIC-IDS2018, as published by the University of New Brunswick's Canadian Institute for Cybersecurity. Our experimental results show that our methods were able to analyze the data within the timeframe during which the data was collected, while also incorporating a larger portion of the data than previous works.

\noindent{\bf Related Work}.
We divide related prior studies into four categories: {\em attack modeling}, {\em attack reconstruction}, {\em alert prioritization}, and {\em intrusion detection}.
In terms of {\em attack modeling}, the problem has been approached using alert correlation \cite{haas2018gac,zhang2019intrusion} and clustering \cite{de2018process}. The present study moves a step further by making sense of alerts through the notion of alert paths, which are more comprehensive than alert correlation because they explicitly model temporal-spatial relationships. This means they may be useful in mapping the attack to a particular model (e.g.,~\cite{CyberKillChainPaper2011,Mandiant,ATTCK}), which may accommodate explicit happens-before dependencies.
Similarly, attack paths (not to be confused with alert paths) have been studied for their usefulness in predicting attack outcomes \cite{chen2007value,hossain2017sleuth,chen2012attack,phillips1998graph,ning2002constructing,ning2003learning,howard2005measuring}. These are distinct from alert paths, which are instead retrospective in nature and based on observed attacker activity rather than potential vulnerabilities.
Attack graphs have also been used to model the potential propagation of attacks through a computer network \cite{ou2006scalable,cinque2020contextual,leitold2016quantifying,lee2018game,hu2018security,Frigault2008}. Like attack paths, attack graphs are focused on static network evaluation or attack prediction, rather than reconstruction of observed attack patterns. 


In terms of {\em attack reconstruction} for multi-step attacks, one work that reconstructs multi-step attacks is MAAC \cite{wang2021maac}. This model assumes a four-stage attack model, where steps of an attack operation can only be assembled if stage 3 alerts are found. This means that false negatives in alert production are more likely to produce false negatives in the model. Although MAAC assembles an alert graph, it only identifies paths of length one, which is likely an attack step within a host rather than across hosts. 
Another reconstruction method is MIF \cite{mao2021mif}, which uses supervised machine learning to reduce graph size and then produces a risk-state graph to track network attacks. Edges are ordered by start times only, which may induce false positives when two paths overlap. The model uses a recursive depth-first-traversal to build paths. 
On LLDoS 2000 (containing 60 hosts), MIF processes one million (upsampled) flows in 3m24s. It processed CICIDS2017 for accuracy but did not give runtimes. 
Yet another reconstruction method is APIN \cite{apin}, which builds an alert graph using raw alerts and extracts alert paths with respect to a particular node using a chronological traversal. It includes complexity analysis and runtimes for the DARPA 1999 intrusion detection evaluation dataset and the CSE-CIC-IDS2018 dataset.



In terms of {\em alert prioritization}, the concept of alert trees benefits from relevant studies of alert prioritization \cite{alsubhi2012fuzmet,ramaki2018systematic,apruzzese2017detection,XuNSS2022-AlertTree}. This process ranks alerts according to their severity or associated risk. These rankings are useful to constructing alert trees because they enable more intuitive tree interpretation. For example, visualizing the colors of nodes in a tree based on the ranking (i.e., prioritization) can help defenders identify hotspots in the network. An innovative method is presented in \cite{XuNSS2022-AlertTree} to achieve better visualization of alert tress.


In terms of {\em intrusion detection}, alert trees depend on input from intrusion detection systems (IDSs), including network-based (NIDS) and host-based (HIDS). 
IDSs have been criticised heavily on account of the base-rate fallacy, poorly-representative environments, limited attack scope, and weak ground truth \cite{axelsson2000base,thakkar2020review,vasilomanolakis2016towards,chou2020data}. 
IDSs are ineffective in practice because of alert volumes, false positives and alert interpretability \cite{mandiant2015numbers,ficke2019analyzing,GabeIEEEMilcom2019,khraisat2019survey}.

\noindent{\bf Paper Outline}.
Section \ref{problem-statement} formalizes the problem.
Section \ref{sec:methods} presents AutoCRAT.
Section \ref{sec:case-study} presents the results of applying AutoCRAT to a dataset. Section \ref{sec:discussion} discusses limitations of AutoCRAT. Section \ref{sec:conclusion} concludes the paper.  
Table \ref{tab:terms} summarizes common terms used in the paper and shows the symbols used to represent them.

\vspace{-1.5em}
\begin{table}[!htbp]
    \centering
    \begin{tabular}{|p{.14\columnwidth}|p{.13\columnwidth}|p{.68\columnwidth}|}\hline
        Term & Symbol & Meaning / Usage \\ \hline\hline
	    Alert/Arc & $\alpha \in A$ (or $\A$) & Event indicating the presence of an attack. By convention, $\A$ denotes the set of all alerts in a dataset, while $A \subseteq \A$.\\\hline
        Alert Graph & $G(\A)$ & The set of vertices, arcs, maps and labels used to model a set of alerts representing network attacks.\\\hline
	    Endpoint pair & $e \in E$ & The ordered pair $(source, destination)$, which is used to group certain sets of arcs.\\\hline
	    Alert Path & $p \in P(\A)$& A sequence of vertices and accompanying arc sets derived from $G(\A)$ with partial happens-before ordering.\\\hline
	    Origin & $v^p_1$ & First node in a path $p$\\\hline
	    Target & $v^p_n$ & Last node in a path $p$ of length $n$\\\hline
	    Alert Tree & $T_{fwd}(\A,\hat{v})$, $T_{bwd}(\A,\hat{v})$ & An aggregation of alert paths with common origin or target, respectively.\\\hline
        Threat Score & $TS(A)$, $ETS(e)$, $PTS(p)$ & Metric used to rank sets of alerts by relative threat to the network. Also used for endpoints and paths, respectively.\\\hline
	    AutoCRAT & & The proposed model for tracking network events. 
	    \\\hline
    \end{tabular}
    \caption{Common terms and symbols used throughout the paper}
    \label{tab:terms}
\end{table}
\vspace{-2.5em}

\section{Problem Statement}\label{problem-statement}

\ignore{
In this section, we formally define the problem addressed in this paper. The terms used are given in Table \ref{tab:terms}. First, we discuss the problem setting and the informal problem statement. 
Then, we define the concepts that will be used to formalize the problem; most notably {\em alert graph}, {\em alert path}, {\em alert tree}. 
Finally, we present the formal research questions.
}

\noindent{\bf Informal Problem Statement}.
Consider an enterprise network, which consists of computer systems and security devices (e.g., NIDSs), and is managed by 
the {\em defender}. Note that the concept of an enterprise network is generally applicable to many types of computer networks, including IoT networks and mobile networks.
Computers on the enterprise network may be the target of cyber attacks from some {\em attacker}, which may come from inside or outside the network. Network traffic is monitored by security devices, which produce alerts when an attack is observed by a security device. 
A successful attacker may conduct secondary attacks (known as {\em lateral movements}) against other computers within the network. 
The term {\em multi-step attack} refers to such a sequence of attacks. 
We aim to develop an understanding of multi-step attacks against a network. This leads to three informal
questions:
(i) What routes could an attacker have taken to get from one computer to another?
(ii) What is the scope of a given attack operation, as represented by an alert or group of alerts?
(iii) How may the attacker have used lateral movements to traverse the network and finally set up a particular attack?
To answer them, we first need to formalize them.

\smallskip

\noindent{\bf Formalization: Alert and Alert Graph / Path / Tree}.
We begin with the input of a stream of alerts, denoted by $\A$, generated by network security / defense devices (e.g., NIDSs). 

\begin{definition}
[Alert]\label{def:alert}
An alert $\alpha$ is generated by a network security device corresponding to a communication between a $source$ computer and a $destination$ computer and can be described as a tuple $\alpha = (source,destination,time,ID)$, where $source$ and $destination$ are the {\em endpoints} of the alert, $time$ represents the timestamp at which the communication begins, and $ID$ is the alert identifier given by the security device (e.g., signature identifier or remote logon type).
\end{definition}


Having defined alerts, we construct more complex objects, namely {\em alert graphs}, {\em alert paths} and {\em alert trees}.
Given a set of alerts, we can construct an {\em alert graph}, 
a labeled multidigraph where  multiple arcs may exist between a pair of nodes (e.g., an attack may make multiple connections before it succeeds, resulting in multiple arcs). Moreover, vertices and arcs can have common labels because alert (i.e., arc) $ID$s belong to a pre-defined set with specific cybersecurity meanings.

\begin{definition}
[Alert Graph]\label{def:alert-graph}
Given a set of alerts $\A$, an alert graph is a labeled multidigraph 
$G(\A) = (\Sigma_V,\Sigma_A,V,A,s,t,\ell_V,\ell_{ID}, \ell_{time})$, where 
vertex $v \in V$ represents a computer on the network, 
arc $a \in A$ represents an alert produced by a security device, 
$\Sigma_V$ denotes the set of computer labels (such as IP addresses), 
$\Sigma_A$ denotes the set of alert labels (i.e., $\alpha$'s $ID$),
$s : A \rightarrow V$ maps arcs to their source vertex (i.e., $\alpha.source$), 
$t : A \rightarrow V$ maps arcs to their target vertex (i.e., $\alpha.destination$),
$\ell_V : V \rightarrow \Sigma_V$ maps vertices to their labels,
$\ell_{ID} : A \rightarrow \Sigma_A$ maps arcs to their alert labels, and 
$\ell_{time} : A \rightarrow \N$ maps alerts to the set of natural numbers, representing the time at which they occurred.
\end{definition}

Given an alert graph $G(\A)$, the concept of {\em alert path} describes a sequence of vertices through which an attack is observed (as indicated by alerts). Since alerts representing a multi-step attack may consist of multiple repeated arcs between a pair of vertices,
a path is associated with a set of arcs where multiple arcs may link the same pair of vertices.
This arc set is further divided into a sequence of smaller arc sets, grouped by the endpoints corresponding to pairs of vertices in the path. Within these arc sets, there must be a valid sequence of arcs, one each from consecutive arc sets, such that they appear in chronological order.

\begin{definition}
[Alert Path]\label{def:alert-path}
Given an alert graph $G(\A)$, an {\em alert} path is defined as $p = (V^p, A^p)$, where $V^p = (v^p_1,\ldots,v^p_n)$ is a sequence of unique vertices in $V$;
$A^p$ is a corresponding set of arcs in $A$, for which there must exist a sequence of arc sets $(A^p_1,\ldots,A^p_{n-1})$ such that $A^p_i = \{\alpha \in A: s(\alpha) = v^p_i \wedge t(\alpha) = v^p_{i+1}\}$, and
$A^p = \bigcup_{i=1}^{n-1}A^p_i$, and
there must exist some sequence $(\alpha_1,\ldots,\alpha_{n-1})$ such that for each $i \in [1,\ldots,n-1]$, $\alpha_i \in A^p_i \wedge i < n-1 \rightarrow \ell_{time}(\alpha_i) < \ell_{time}(\alpha_{i+1})$.

\end{definition}

Given an alert path $p$ with $V^p = (v^p_1,\ldots,v^p_n)$, we say $v^p_1$ is
the {\em origin} and $v^p_n$ is the {\em target}.
Given $\A$, 
let $P(\A)$ denote the set of all alert paths in $G(\A)$. 
A path $p \in P(\A)$ may be uniquely identified by $V^p$.


An alert tree represents a set of alert paths $P(A)$ (where $A \subseteq \A$) with a common {\em reference vertex}, which is either the origin or the target, but not both. Trees with a common origin are called {\em forward trees} and trees with a common target are called {\em backward trees}, because of the 
happens-before relationship of alerts in the tree. Because the graph $G(\A)$ is a labeled multidigraph, it is possible that cyclical subgraphs will result in sets of paths $P(A)$ such that they cannot naturally be combined to form trees.
For example, the graph constructed from the set of paths $P = \{(a,b,c),(a,c,b)\}$ contains a cycle $(b,c,b)$ and is therefore not a tree. However, if we manipulate the set of paths by changing duplicate nodes into unique nodes with the same identifier (in $\Sigma_V$), we can artificially prevent this loop formation. Consider the previous example. We can change the second path to $(a,c',b')$ such that $\ell_V(c') = \ell_V(c)$ and $\ell_V(b') = \ell_V(b)$. This results in a tree with arcs $(a,b),(b,c),(a,c'),(c',b')$ such that the modified graph forms a tree, which can be used to extract the relevant computer identifiers. The advantage of the tree structure over an arbitrary graph is that
one can visualize the {\em temporal dependence} of the arcs based on their height within the tree. Since temporal dependence is an important feature of alert paths, it is natural to visualize sets of alert paths using alert trees rather than alert graphs.

\ignore{
\begin{definition}
[Alert Tree Node]\label{def:node}
Given an alert graph $G = (V,E)$, an alert tree node $n = (name,parent,children,color)$; where $n.name = v.name$ for some $v \in V$; $n.parent$ is either a single alert tree node or, in the case of the root node, $\emptyset$; $n.children = \{n'_1,\ldots,n'_m\}$, a set of alert tree nodes such that $n' \in n.children \rightarrow n'.parent = n \wedge \forall n'' \in n.children, n''.name = n'.name \rightarrow n'' = n'$; and $n.color$ is discussed below, in the context of the alert tree as a whole. 
\end{definition}

Each alert tree node $n$ can be said to have a {\em genealogy} (denoted $Gen(n)$), a sequence of alert tree nodes where the first member has no parent, each successive member is a child of the previous member, and n ends the sequence. Genealogy is formally defined in Definition \ref{def:genealogy}

\begin{definition}
[Genealogy]\label{def:genealogy}
The genealogy of an alert tree node $Gen(n) = Gen(n.parent) \cup (n)$, and $Gen(\emptyset) = ()$. The inverse of a genealogy $Gen(n)^{-1} = (n) \cup Gen(n.parent)$ and $Gen(\emptyset)^{-1} = ()$.
\end{definition}

Alert tree nodes are used to construct alert trees, which come in two forms: forward and backward. Forward trees model attacks that spread from a single attacker to one or more targets. Backward trees model attacks from one or more attackers to a single target. These are defined in Equations \eqref{def:tree-fwd} and \eqref{def:tree-bwd}, respectively.
}

\begin{definition}
[Alert Tree]\label{def:tree}
Given an alert graph $G(\A)$, a reference vertex $\hat{v} \in V$, and a set of paths $P' = \{p \in P(\A) : v^p_1 = \hat{v}\}$, 
a forward alert tree is a labeled digraph denoted $T_{fwd}(\A,\hat{v}) = \{\Sigma_{V,T},V_T,A_T,\ell_{V,T}\}$, where 
$\Sigma_{V,T} = V$, 
$V_T = V \times P'$,
$A_T = (v,v') \in V_T$ such that $\exists p \in P', i \in [1,\ldots,|V^p|]$ for which $v = v^p_i \wedge v' = v^p_{i+1}$,
and $\ell_{V,T} : V_T \rightarrow \Sigma_{V,T}$ maps vertices to their corresponding nodes in $V$. 
A {\em forward alert tree} is rooted at reference vertex $\hat{v}$ and represents paths $p \in P'$, such that every vertex in $p$ has a corresponding vertex $v \in V_T$ for which all of the ascendants of $v$ (denoted $asc(v)$) are labeled with the vertices in the path. 
Specifically, $\forall i \in [1,\ldots, |asc(v)|], \ell_{V,T}(asc(v)_i) = v^p_i$.
A {\em backward alert tree} $T_{bwd}$ is a reversed forward tree in the following sense: reference vertex $\hat{v}$ is the target rather than the origin of all of the corresponding alert paths, while the vertex identifiers must match vertex descendants rather than ascendants.
\end{definition}




Equipped with the preceding definitions, we can now formalize the afore-mentioned informal questions as the following research questions (RQs).


\begin{itemize}
\item 
{\bf RQ1}: This question asks for the set of all alert paths with a specified origin and target. Formally, given a set of alerts $\A$, a known attack origin $v_{origin}$, and a known attack target $v_{target}$, produce the set of attack paths $P' = \{p \in P(\A) : v^p_1 = v_{origin} \wedge v^p_{|V^p|} = v_{target}\}$.

\item {\bf RQ2}: 
This question asks for the depth and breadth of access that an attacker achieved corresponding to an alert (or attack). 
Formally, given a set of alerts $\A$ and a known attack origin $v_{origin}$, produce the forward alert tree $T_{fwd}(\A,v_{origin})$.

\item {\bf RQ3}: 
This question asks for the set of multi-stage attacks that may have led to the compromise of a computer. 
Formally, given a set of alerts $\A$ and a known attack target $v_{target}$, produce the backward alert tree $T_{bwd}(\A,v_{target})$.

\ignore{
\item What is the most efficient way to achieve each of the above goals? Of course, this is a significant optimization problem. However, we can compare the algorithmic complexity and practical runtime efficiency of the proposed algorithms to those of existing models. This leads to:

{\bf RQ4}: {\em Analyze the asymptotic runtime and storage complexities of the proposed methods with respect to their input alerts $\A$.}
}
\end{itemize}



\ignore{
\subsection{Threat Model}
\label{sec:threat-model}

We consider an attacker that may have access to computers inside or outside  the network. The attacker can accomplish this through various means, such as social engineering attacks, the specifics of which are outside the scope of the present study. Specifically, we consider that any attack detectable by a network security device is within the present threat model. Once the attacker has control over a computer in the network, the attacker can leverage the compromised computer to conduct lateral movements within the network. Tracking these movements is the primary motivation for this work, and is a key component of the threat model.

\noindent{\bf Assumptions}.
We make the following assumptions.
First, the attacker cannot break the security devices; otherwise, the attacker can selectively turn off alerts at will. We stress that this does not mean the defense devices are perfect; to the contrary, these devices are limited because they may have (for example) a high false-positive rate, which actually partly motivates the present study in that threat score minimizes the impact of a small amount of false-positives.
Second, the attacker cannot manipulate the analysis system implementing our algorithms; otherwise, the attacker can manipulate the input to, or output of, the analysis system at will. The preceding two assumptions collectively imply that we can use standard cryptography (e.g., message authentication code) to protect the integrity of streams of alerts when transmitted from security devices to the analysis system.
Third, the streams of alerts can always be transmitted from security devices to the analysis system in a timely manner, which means that the attacker cannot wage denial-of-service attacks. 
We are confident that the preceding three assumptions are orthogonal to the focus of the present paper, and note that they are achievable in practice in most cases as, to the best of our knowledge, no incidents violate these three assumptions. 

We also make an assumption about the behavior of the network as a whole, in order to clearly convey ideas while minimizing the use of notations. Specifically, we assume that $V$ is static, meaning that the set of computers on a network is fixed. This may be a weak assumption, as our methods can easily be extended to accommodate time-dependent sets of computers (e.g., by adding a custom alert $\alpha$ to indicate that a computer's IP address has moved from $\alpha.source$ to $\alpha.destination$).
}

\section{The AutoCRAT System}
\label{sec:methods}
\begin{figure}[!htbp]
    \centering
\includegraphics[width=.9\linewidth]{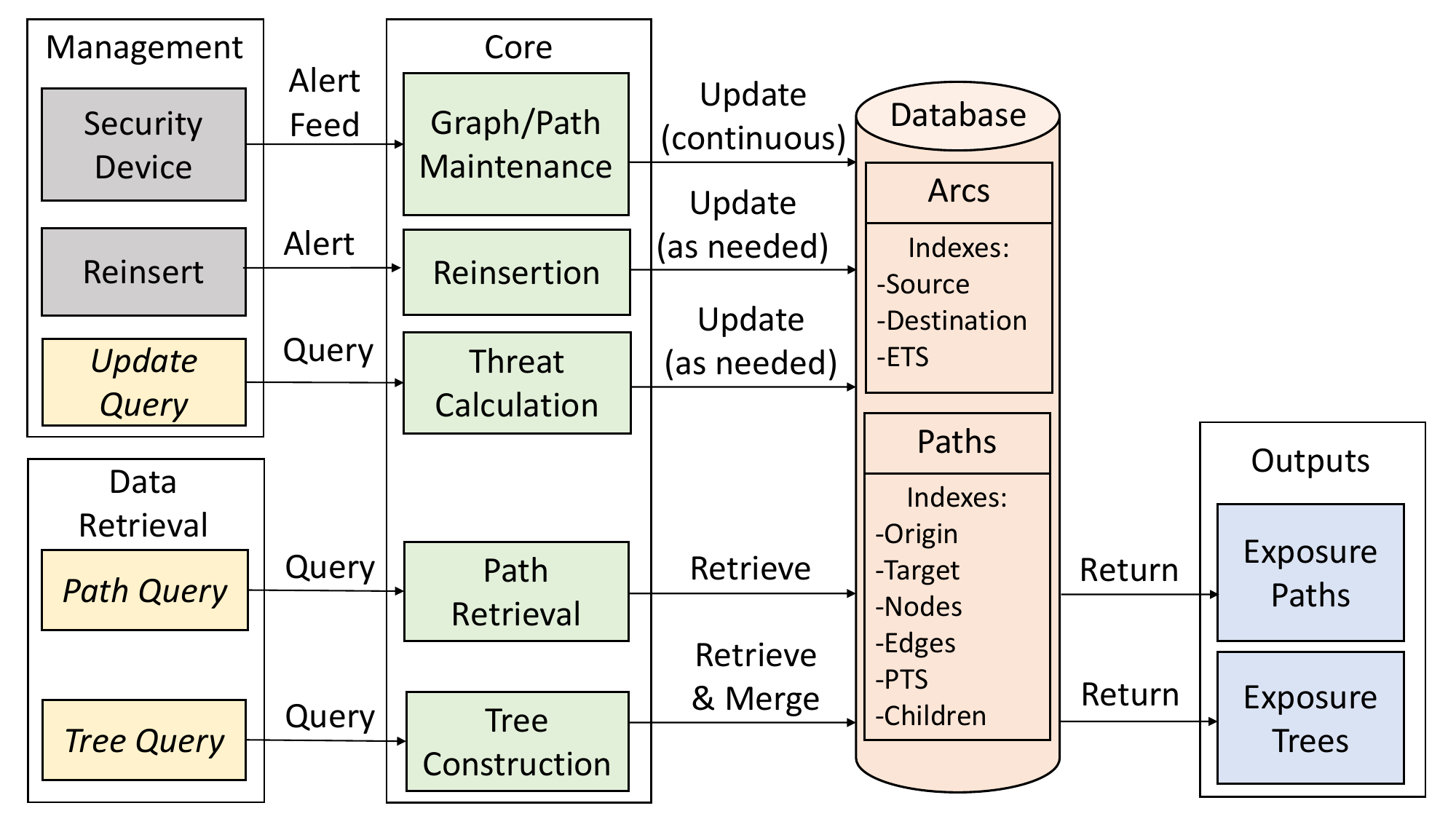}
\vspace{-1em}
    \caption{Architecture of the AutoCRAT system}
    \label{fig:architecture}
\end{figure}

Here we propose the AutoCRAT system to address the RQs mentioned above. Figure \ref{fig:architecture} highlights its architecture, which has two main components: {\em database} and {\em core} functions. The system receives inputs in the form of {\em management} and {\em data retrieval}, which make calls to the core functions to change or retrieve data from the database, respectively. This section discusses the {\em database} and {\em core} functions of AutoCRAT.

\vspace{-1.5em}
\vspace{-2em}

\subsection{Database}
The database stores the arcs (i.e., alerts) and paths derived from $G$ as follows. 
First, the arc collection of the database stores pairs of arc endpoints. 
Arcs are stored as annotations to the arc endpoints. This enables more efficient retrieval of alerts, which are always considered in the context of the other alerts with the same endpoints. 
Second, alert paths are stored in their own collection. Each path $p \in P(\A)$ contains a list of nodes, the corresponding node pairs (to facilitate retrieval via pairwise indexing), the path's PTS, and a list of child vertices (denoted $Children^p$) that have been used to produce lengthened copies of the path, $\{v \in V : (\exists p' \in P(\A), \forall v^p_i \in V^p) v^p_i = v^{p'}_i \wedge |V^{p'}| = |V^p| + 1 \wedge v =  v^{p'}_{|V^{p'}|}\}$. This list is used to facilitate path validation, as discussed later in this section. Every time a path is lengthened, the new path is added as its own object in the database, in order to facilitate indexing as described below.


\noindent{\bf Indexing}. Each collection is indexed so that its objects may be efficiently found and retrieved from the database. Our approach uses multiple types of indices, including individual indexes (on a single field), compound indexes (on two or more fields) and multikey indexes (on a set or sequence). 
Some of these index types are only available in certain kinds of databases, 
such as MongoDB, which is used in our implementation. This limits the interoperability of our approach, or risks compromising the efficiency of accessing some objects in the database. 

The arc collection is indexed by arc endpoints (each individually as well as together in a compound index), and by the arc's threat score (discussed later in this section). The path collection has individual indices on the origin, the target, and the path's threat score (also discussed later). It also has multikey indexes on the vertex sequence and the arc endpoint sequence. Additionally, the path collection has one compound index, covering the target and the list of child vertices (as a multikey index). 

Duplicate paths are prevented by validating new arcs against this list of children in the path's so-called parent (i.e., $p' \in P$ such that $V^{p'} = V^p - (v^p_{|V^p|}) \wedge v^p_{|V^p|} \in Children^{p'}$. 
Subpaths are stored as their own objects because this allows them to be more efficiently indexed. Specifically, we index the path's origin and target, because multikey indexes do not preserve order. This means that, if a multikey index was used on the path nodes, then an attempt to retrieve paths with a certain origin and target would need to parse all paths containing both the origin and target in any position, then trim the path appropriately. In this case, the runtime optimization of multikey indexing of path vertices trades off with storage complexity (i.e., storing more paths). 

On the other hand, it is efficient to use a multikey index to index to children of a particular path, because queries that search for children need only find a single child (i.e., the $destination$ of the alert being inserted). This means that the ordering of children in a multikey index is unimportant.

\subsection{Core}
The core has five modules: {\em alert graph / path maintenance}, {\em reinsertion}, {\em threat calculation}, {\em path retrieval}, and {\em tree construction}. These modules are invoked by the respective management services.

\vspace{-1.5em}
\begin{algorithm}[!htbp]
{\scriptsize
    \caption{Alert Graph / Path Maintenance}\label{alg:maintenance}
    \textbf{Input:} $\alpha,G(\A) = (\Sigma_V,\Sigma_A,V,A,s,t,\ell_V,\ell_{ID}, \ell_{time}),P,Children^{p \in P}$\\
    \textbf{Output:} $G(\A \cup \{\alpha\}), P, Children^{p \in P}$ 
    \begin{algorithmic}[1]
    \If{$\not\exists \alpha' \in A : s(\alpha) = s(\alpha') \wedge t(\alpha) = t(\alpha')$}
        \State $V^{\hat{p}} \gets (s(\alpha),t(\alpha))$
        \Comment{Create $\hat{p}$ via $V^{\hat{p}}$}
        \State $Children^{\hat{p}} \gets \emptyset$
        \State $P \gets P \cup \{\hat{p}\}$
    \EndIf
    \State $A \gets A \cup \{\alpha\}$
    \State $V \gets V \cup \{s(\alpha),t(\alpha)\}$
    \State $Candidate\_Paths \gets \{p \in P : V^p_{|V^p|} = s(\alpha)\}$ \Comment{Find paths to lengthen using ``end'' index}
    \For{$c \in Candidate\_Paths$}
        \If{$t(\alpha) \not\in V^{c}$} \Comment{Prohibit cycles}
            \If{$t(\alpha) \not\in Children^{c}$  } \Comment{Prohibit identical twins}
                \State $p' \gets \{V^c,A^c\}$
                \State $V^{p'} \gets V^{p'} \cup (t(\alpha))$
                \State $A^{p'} \gets A^{p'} \cup \{\alpha\}$
                \State $Children^{p'} \gets \emptyset$
                \State $Children^c \gets Children^c \cup \{t(\alpha)\}$
                \State $P \gets P \cup \{p'\}$
            \EndIf
        \EndIf
    \EndFor
    \State {\bf return} $G=(V,A), P, Children^{p \in P}$
    \end{algorithmic}
}
\end{algorithm}
\vspace{-2em}

\noindent{\bf Core Module 1: Alert Graph / Path Maintenance}.
As new security events arrive, AutoCRAT incorporates them into the relevant databases according to Algorithm \ref{alg:maintenance}, which proceeds as follows: (i) It first checks if the new arc includes existing endpoints. If so, it adds the arc to the endpoint object as an annotation; otherwise, it inserts a new endpoint object annotated with the arc (Lines 1-13). (ii) It queries the database to find paths which end at the source of the event (Line 14). (iii) The algorithm copies the paths found, appending the alert's $destination$ onto the copies. If the new paths already exist in the database or are cyclical, they are discarded. The original paths are annotated with a list of children to facilitate this check on future inserts, and the new paths are inserted (Lines 15-27).



In the worst-case scenario, every insertion of a new alert $\alpha$ will add a new endpoint object and path, and lengthen the set of existing paths that terminate at the source of the inserted arc. 
This will give us a storage complexity in $\mathcal{O}(|E| + |A| + |P|)$. 
Now suppose one inserts the first arc, with endpoints $(v_1,v_2)$, into a database. This will produce a single path: $(v_1,v_2)$. A second insertion of $(v_2,v_3)$ adds a new path and lengthens the existing path, resulting in three paths: $(v_1,v_2)$, $(v_2,v_3)$, and $(v_1,v_2,v_3)$. Clearly, each arc adds at most one new endpoint object (exactly one in this case), leading to $|E|$ or $\mathcal{O}(|\A|)$. This means that each new arc adds up to $|E|$ new paths of length $|E|$ or less, leading to $|P| \in \mathcal{O}(|\A|^2)$. This is demonstrated in Table \ref{tab:runtime-worst-case-insert}. 

\vspace{-1.5em}
\begin{table}[!htbp]
    \centering
    \begin{tabular}{|c|c|c|c|c|c|}\hline
        Arcs           & 1 & 2 & 3 & 4  & $|\A|$ \\\hline
        Endpoint pairs & 1 & 2 & 3 & 4  & $|\A|$ \\\hline
        Paths          & 1 & 3 & 6 & 10 & $\sum_{i=1}^{|\A|}(i)$\\\hline
    \end{tabular}
    \caption{The worst-case storage complexity after $i$ arc insertions.
    As each new arc $\alpha_i$ can add as many as $i$ paths, the worst-case number of paths is $|P| = \frac{1}{2}|\A|^2 + \frac{1}{2}|\A|$ or $\mathcal{O}(|\A|^2)$.}
    \label{tab:runtime-worst-case-insert}
\end{table}
\vspace{-2em}

Under the same worst-case scenario,
the insert algorithm must access every path in the database, $|P|$, copy each, adding one endpoint, and insert the copies, plus one new path with two nodes. Then the worst-case runtime (in terms of the number of accesses to the database) is $2\cdot|P| + 1$ or $\mathcal{O}(|P|)$.
\smallskip

\noindent{\bf Core Module 2: Reinsertion}.
In some cases, the defender may need to find paths that were unavailable at the time of data collection (e.g., if the NIDS produces a false negative which is later corrected). Since alerts are otherwise inserted chronologically,
we need another algorithm to perform such retroactive insertions (``reinsert''). For this purpose we propose Algorithm \ref{alg:reinsert}.

\begin{algorithm}[!htbp]
\footnotesize
\caption{Alert Reinsertion}\label{alg:reinsert}
\textbf{Input:} $\alpha,G(\A),P$\\
\textbf{Output:} $G(\A \cup \{\alpha\}), P$ 
\begin{algorithmic}[1]
    \State $P^{pre} \gets \{p \in P : V^p_{|V^p|} = s(\alpha)\}$
    \State $P^{pre} \gets P^{pre} \setminus \{p \in P^{pre} : t(\alpha) \in V^p\}$
    \State $P^{post} \gets \{p \in P : V^p_{1} = t(\alpha)\}$
    \State $P^{post} \gets P^{post} \setminus \{p \in P^{post} : s(\alpha) \in V^p\}$
    \State $V \gets V \cup \{s(\alpha),t(\alpha)\}$
    \State $A \gets A \cup \{\alpha\}$
    \State $V^{\hat{p}} \gets (s(\alpha),t(\alpha))$
    \Comment{Create $\hat{p}$ via $V^{\hat{p}}$}
    \State $P \gets P \cup \{\hat{p}\}$
    \For{$p \in P^{pre}$}
        \For{$p' \in P^{post}$}
            \If{$V^p \cap V^{p'} = \emptyset$}
                \Comment{Prohibit Cycles}
                \State $V^{New\_Path} \gets V^p \cup V^{p'}$
                \Comment{Create $New\_Path$ via $V^{New\_Path}$}
                \If{$New\_Path$ is valid given $A$}
                    \State $P \gets P \cup \{New\_Path\}$
                \EndIf
            \EndIf
        \EndFor
    \EndFor
    \State {\bf return} $G,P$
\end{algorithmic}
\end{algorithm}

The storage complexity is the same as for the regular maintenance function, $\mathcal{O}(|\A|^2)$.
The asymptotic runtime is dominated by $|p^{pre}| \cdot |p^{post}|$.  
Because cycles about $\alpha$ are removed (lines 2,4), $P^{pre} \cap P^{post} = \emptyset$. This is bounded by $(\delta \cdot |P|) \cdot ((1-\delta) \cdot |P|)$, where $\delta < 1$ is the proportion of the larger of the two path sets relative to $|P|$. This simplifies to $(\delta - \delta^2)\cdot|P|^2$, for which the maximum value of the first term (when $\delta = \frac{1}{2}$) is $\frac{1}{4}$. 
The worst-case runtime is $\frac{1}{4}|P|^2$ or $\mathcal{O}(|P|^2)$.

\smallskip

\noindent{\bf Core Module 3: Threat Calculation}.
We propose measuring a given set of alerts $A \subset \A$ according to its threat score (TS), which can also be used during alert visualization. For example, nodes in a graph may be colored from black to red based on their TS in ascending order, to guide the viewer to network hotspots. We use this approach in our examples to follow.

There can be many definitions for TS, and identifying the ``best'' definition is orthogonal to the focus of the present paper. Since any ``good'' definitions can be incorporated into AutoCRAT in a plug-and-play fashion, we will use one example definition to demonstrate the idea. In this example definition of TS, we define it as the geometric mean of the number and diversity of alerts in a set, in order to estimate the risk posed by the associated attacks. This definition can be naturally extended to specify the TS of a pair of endpoints (Endpoint Threat Score --- ETS) or a path (Path Threat Score --- PTS). Specifically, we have:
\begin{definition}[Threat Score (TS)]
The TS of a set of alerts is defined as:
\begin{equation}
    \label{eq:ts}
    TS(A) = \sqrt{|\{\sigma \in \Sigma_A:(\exists \alpha \in A),\sigma = \ell_{ID}(\alpha)\}| \cdot |\{\alpha \in A\}|}.
\end{equation}
The ETS of an endpoint pair $(source,destination)$ can be calculated as $TS(\{\alpha \in \A : s(\alpha) = source \wedge t(\alpha) = destination\})$.
The PTS of a path $p$ can be calculated as $TS(A_p)$.
\end{definition}

In order to identify the endpoints and paths in the database with the highest threat, AutoCRAT must periodically update ETS and PTS for endpoints and paths, respectively. In order to do this, the defender submits an update query on demand. The update function is simple: first, it calculates ETS for every endpoint object, then it calculates PTS for every path. To reduce the number of accesses to the database, a copy of each endpoint object's arc annotations is cached during the ETS calculations, in order to facilitate the PTS calculations. 


The runtime of this approach is based on the following: (i) calculating ETS for all endpoints requires the parsing of each arc (which each belong to exactly one endpoint annotation), meaning the runtime of updating is in $\Omega(|\A|)$; (ii) calculating PTS for all paths requires the parsing of each endpoint object a number of times equal to the number of paths in which it appears. This is analyzed in Table \ref{tab:runtime-worst-case-edge-count-in-paths}, which shows that the worst case number of appearances of an endpoint object in the set of paths is in $\mathcal{O}(|E^2|)$. Since there are $|E|$ endpoint objects, this results in a worst-case combined runtime in $\mathcal{O}(|\A| + |E|^3$). 

\vspace{-1.5em}
\begin{table}[!htbp]
    \centering
    \begin{tabular}{|c||c|c|c|c|c|c|}\hline
        Endpoint $i$ & Insertion 1  & 2 & 3 & 4 & 5 & $|E|$\\\hline\hline
        1                           & 1 & 2 & 3 & 4 & 5 & $|E|$ \\\hline
        2                           & 0 & 2 & 4 & 6 & 8 & $2\cdot (|E|-1)$ \\\hline
        3                           & 0 & 0 & 3 & 6 & 9 & $3\cdot (|E|-2)$ \\\hline
        4                           & 0 & 0 & 0 & 4 & 8 & $4\cdot (|E|-3)$ \\\hline
        5                           & 0 & 0 & 0 & 0 & 5 & $5\cdot (|E|-4)$ \\\hline
        \hline$|E|$                 & 0 & 0 & 0 & 0 & 0 & $|E|            $\\\hline
        \hline$i$                   & 0 & 0 & 0 & 0 & 0 & $i \cdot (|E|-i+1)$\\\hline
    \end{tabular}
    \caption{The worst-case number of paths that contain a particular endpoint.
    The $i^{th}$ endpoint can only belong to 
    $i \cdot (|E|-i+1) = i\cdot |E| - i^2 + i$
    paths. Then the worst-case is in
    $\mathcal{O}(|E|^2)$
    , and the worst-case number of endpoint objects across the set of all paths is
    $\sum_{i=1}^{|E|}i\cdot( |E| - i + 1) = $
    $\frac{1}{6} |E|(|E|+1)(|E|+2)$ or $\mathcal{O}(|E|^3)$.}
    \label{tab:runtime-worst-case-edge-count-in-paths}
\end{table}

\vspace{-2em}

\ignore{
\begin{definition}
[Edge Threat Score (ETS)]\label{def:ets}
 Given a set of security events traversing an edge $e$ in a network graph, $ETS(e)$ is the geometric mean of the diversity of alerts along the edge and the quantity of alerts along the edge. Diversity of an edge is calculated in Equation \eqref{eq:diversity}, and ETS is calculated in Equation \eqref{eq:ets}.

\begin{equation}\label{eq:diversity}
    Diversity(e) = |\{\alpha .ID | \alpha \in e.events\}|
\end{equation}

\begin{equation}\label{eq:ets}
    ETS(e) = \sqrt{Diversity(e) \cdot |e.events|}
\end{equation}
\end{definition}
}

\begin{figure}[!htbp]
    \centering
\includegraphics[width=.3\textwidth]{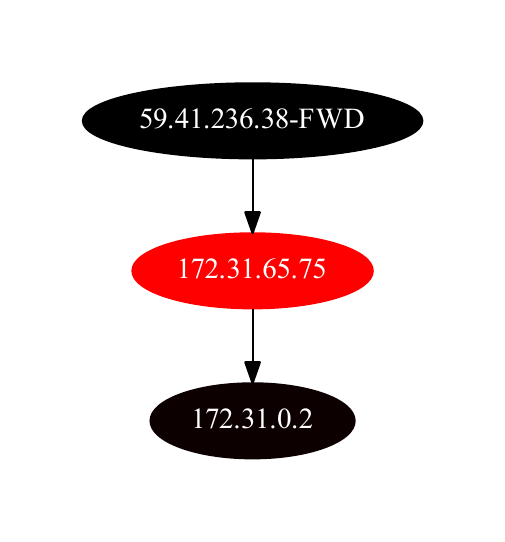}
\vspace{-2em}
\caption{Example alert tree coloring, where the root is black, the child is red (ETS 179.10), and the grandchild is very nearly black (color code 0x0D0000 and ETS 10.49). 
    }
    \label{fig:color}
\end{figure}

ETS is also used in the visualization of attack paths. Specifically, the color of an alert tree node $n$ represents the normalized ETS of $(parent(n),n)$ for forward trees or $(n,parent(n))$ for backward trees. 
Vertex colors range from red to black, where the vertex in the tree with the highest ETS is red and the root is black (along with any other vertices with $ETS = 1$). In RGB (i.e., hexadecimal) notation, colors range from 0x000000 (black) to 0xFF0000 (red), such that colors may be compared ordinally to mimic the comparison of ETS. This is demonstrated in Figure \ref{fig:color}, showing the value of the ETS measurement in presenting salient information to the defender. 

\vspace{-1.5em}
\vspace{-2em}

\ignore{
\subsection{{\color{red}User Interface}}

{\color{blue}
This component is responsible for defender-AutoCRAT interactions. Specifically, a defender can make queries to AutoCRAT, such as those specified by the RQ1-RQ3 mentioned above. Nevertheless, the following kinds of queries may be made by a defender.
(i) A defender asks AutoCRAT to update threat scores of database objects.
(ii) A defender asks AutoCRAT to retrieve alert paths from the database, where a query specifies the path's origin or target or allows threat score heuristics to automatically select them.  
(iii) A defender asks AutoCRAT to reconstruct alert trees from stored alert paths, where a query specifies the root of the tree and the direction in which to build the tree (i.e., forward or backward).
}
}


\smallskip

\noindent{\bf Core Module 4: Path Retrieval}.
Path retrieval requires that the graph and path databases are properly maintained. 
The path retrieval process then involves a simple query to the database, which stores each path individually. This query leverages the $(source,destination)$ compound index, which efficiently retrieves the appropriate paths from the database. The retrieval algorithm executes with a runtime efficiency of $\mathcal{O}(|P|)$, where $P \subseteq P(\A)$ is the set of paths to be retrieved.

\smallskip

\noindent{\bf Core Module 5: Tree Reconstruction}.
Tree reconstruction is to reassemble paths into a tree which represents either the attack surface exposed to an attacker or the attack vectors exposing a target, depending on the type of query. In either case, the user must specify a node to act as the root of the tree and the tree's direction, and the module handles the rest. 
The tree reconstruction algorithm takes as input a reference node to act as the tree's root, a direction (i.e., forward or backward), $G$, and $P$. It begins by selecting all paths which have the reference node as their origin (for forward trees) or target (for backward trees), using the approach described above. It then parses each path, adding each edge as a node of the tree if it was not already added from a previous path. 


The worst-case scenario for the tree reconstruction algorithm is the same as the insert algorithm as discussed above, in which each subsequent alert produces a new endpoint which extends that of the previous alert. Specifically, we have at most $|E|$ paths rooted at a given node, and the maximum path length is $|E|$. This results in a final worst-case runtime of $\mathcal{O}(|E|^2)$.

\section{Case Study}
\label{sec:case-study}

Our experiments were run using Ubuntu 20.04 with 192GB of RAM, 2 cores of an Intel Xeon Gold 6242 CPU @2.80 GHz, and a 200GB HDD. These resources were shared among AutoCRAT functions and the corresponding MongoDB database, which was installed on the same computer to eliminate variability imposed by network conditions.

\subsection{Dataset}

The dataset was published by the University of New Brunswick, and is referred to as CSE-CIC-IDS2018 \cite{sharafaldin2018toward}. It contains data collected over the course of 9 days, during which multiple distinct attack scenarios were executed against the network. The environment was connected to the internet during the experiments, thus real-world attacks can also be observed in the data. We preprocessed the packet capture (PCAP) files in the dataset using Suricata 4.0 \cite{suricata} with the corresponding Emerging Threats signature set \cite{emerging_threats_2019}, to produce a set of 3,323,426 alerts, of which 19,921 were strictly internal to the target network as defined by the dataset authors. We converted the alerts into JSON objects to conform to AutoCRAT's expected format, and sorted them chronologically before feeding them into the database.

\subsection{Experimental Results}

Database construction for CSE-CIC-IDS2018 took 35h14m07s, resulting in 1,053,710 edges and 3,591,217 paths.
Efficiency and accuracy (in terms of graph coverage, which may impose false negatives) are analyzed relative to an existing model, APIN \cite{apin}. 
Path selection for APIN was done using its relevant heuristics. The comparison is shown in Table \ref{tab:comparison-orig}.

\vspace{-1.5em}
\begin{table}[!htbp]
    \centering
    \begin{tabular}{|c||c|c|}\hline
                                & \cite{apin}   & \textbf{AutoCRAT}   \\\hline\hline
         Build DB               & 29m43s        & 13h42m41s         \\\hline
         Rank Objects*          & 49s           & 1h00m29s          \\\hline
         Top 100 paths          & 52s†          & \textbf{32ms}              \\\hline
         Top 20 trees           & 52s†          & 2m42s†            \\\hline
         Coverage (nodes)       & 99.6\%        & \textbf{100\%}             \\\hline
         Coverage (events)      & 3.4\%         & \textbf{100\%}              \\\hline
         DB size                & 637 MB        & 1.1 GB           \\\hline
    \end{tabular}
    \caption{Comparison of the proposed AutoCRAT model with an existing model \cite{apin}. Query runtimes are the average of 10 runs. Improvements shown in bold.
    *\cite{apin} only ranks nodes, while AutoCRAT scores endpoints and paths. 
    †The models do not directly rank these objects, so they are selected based on applicable node and edge rankings.
    These inherited rankings may not be accurate with respect to other metrics but offer a reasonable baseline.
    }
    \label{tab:comparison-orig}
\end{table}
\vspace{-2em}

Note that \cite{apin} sacrifices coverage in order to improve runtime. This is necessary because its tree retrieval time suffers extraordinary slowdown in the presence of highly connected nodes. Even though \cite{apin} only excludes .4\% of the vertices in the graph, 96.6\% of the arcs in the graph are adjacent to these vertices, and are effectively blacklisted from tree reconstruction, meaning that the set of paths (and trees) that can be reconstructed is incomplete. In AutoCRAT, the connectedness slowdown problem is solved by shifting the bulk of the work into the pre-processing stage. This results in a much faster path retrieval time and full graph coverage at the cost of maintenance time. However, this pre-processing time remains feasible in practice since 9 days of data (constituting approximately 164 hours of activity) are processed in under 14 hours. Graph coverage is important because low coverage induces false negatives in the path and tree reconstruction. This means that the model in \cite{apin} is vulnerable to DoS attacks, which may allow an attacker to conceal their attack paths by creating so much traffic around a key node that the algorithm removes it from the analysis entirely.

In order to compare the two models under comparable conditions, we ran both models again using a reduction of the original dataset, which filtered nodes outside the network as well as edges which crossed the border of the network. This resulted in only 1,994 edges and 3,019 paths, as shown in Table \ref{tab:comparison-reduced}.
This resulted in a coverage of 0.6\% for both nodes and paths, much closer to that of \cite{apin} (i.e., roughly 1/6). Given the intuition that internal nodes are far more relevant to defenders, we do not consider the loss of node coverage important in this case (although some of the nodes filtered by \cite{apin} were in fact internal nodes). This reduction greatly improved the performance of AutoCRAT, resulting in runtimes and storage efficiency that far outperformed \cite{apin}. 


\begin{table}[!htbp]
    \centering
    \begin{tabular}{|c||c|c|}\hline
                                & \cite{apin}-internal      & \textbf{AutoCRAT}-internal  \\\hline\hline
         Build DB               & 9s                        & 35s\\\hline
         Rank Objects*          & 0.28s                     & 5s\\\hline
         Top 100 paths          & 3s†                       & \textbf{23ms}\\\hline
         Top 20 trees           & 3s†                       & \textbf{1.97s}†\\\hline
         Coverage (nodes)       & 0.6\%                     & 0.6\%\\\hline
         Coverage (events)      & 0.6\%                     & 0.6\% \\\hline
         DB size                & 2.9 MB                    & \textbf{2.4 MB}\\\hline
    \end{tabular}
    \caption{Comparison of the proposed AutoCRAT model with an existing model \cite{apin}, using only the internal events. Query runtimes are the average of 10 runs. Improvements shown in bold.
    *\cite{apin} only ranks nodes, while AutoCRAT ranks endpoints and paths. 
    †The models do not directly rank these objects, so they are selected based on applicable node or endpoint rankings.
    These inherited rankings may not be consistent with respect to other metrics but offer a reasonable baseline.
    }
    \label{tab:comparison-reduced}
\end{table}

The above discussion demonstrates trade-offs between pre- and post- processing times and between processing time and accuracy. 
\begin{insight}\label{ins:tradeoff-pre-post-time}
To attain the same accuracy, AutoCRAT (relative to \cite{apin}) front-loads its processing time into building and maintaining paths so that when it comes time to retrieve them, it can do so more quickly. 
For both models, reducing the volume of data processed improves both processing time and database size, but sacrifices accuracy (as measured by coverage).
\end{insight}

\vspace{-3em}
\begin{figure}[!htbp]
    \centering
\includegraphics[width=.8\columnwidth]{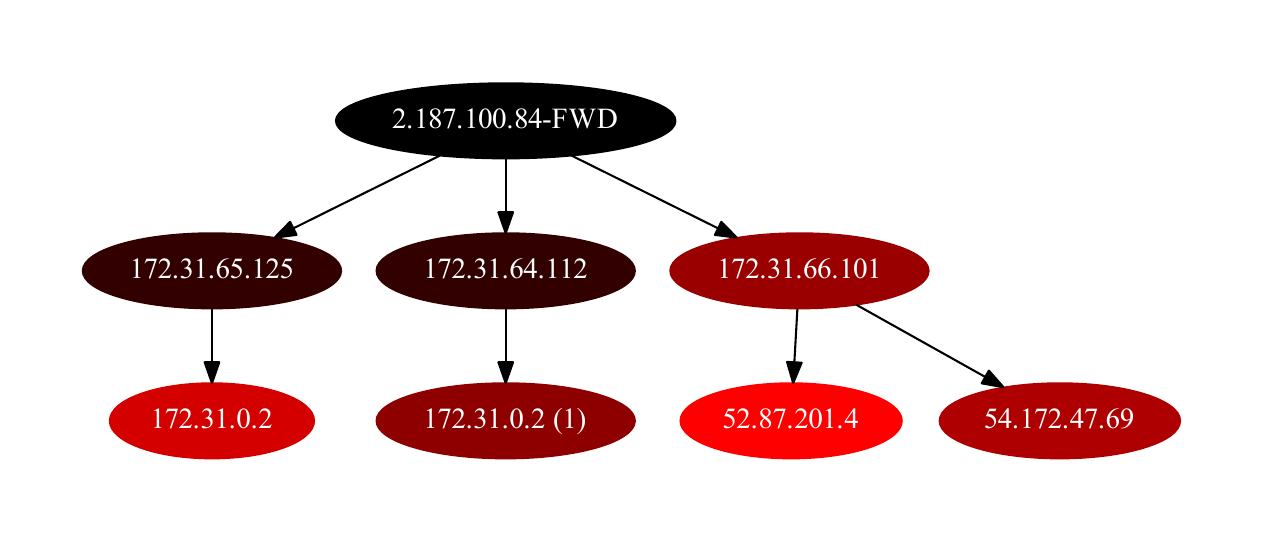}
\vspace{-2em}
    \caption{A forward tree containing eight vertices, colored red to black based on normalized ETS, descending. Of the four leaves (i.e., path targets), the reddest vertex, which represents the endpoints $(172.31.66.101,52.87.201.4)$ scored an ETS of 5.92 with 35 alerts sharing a single $ID$.}
    \label{fig:fwd-tree}
\end{figure}
\vspace{-1.5em}

\begin{figure}[!htbp]
    \centering
\includegraphics[width=.9\columnwidth]{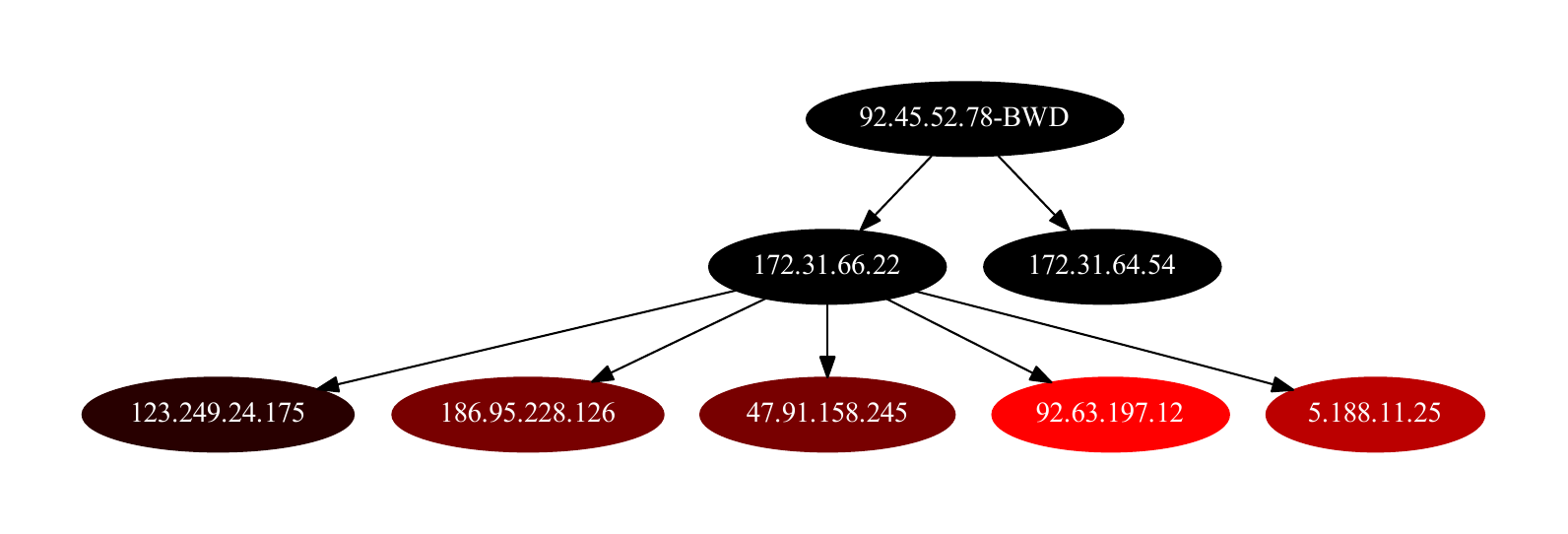}
\vspace{-1em}
    \caption{A backward tree containing eight verticess, colored red to black based on normalized ETS, descending. Of the six leaves (i.e., path origins), the reddest vertex, which represents the endpoints $(92.63.197.12,172.31.66.22)$ scored an ETS of 7.35 with 54 alerts sharing a single $ID$.}
    \label{fig:bwd-tree}
\end{figure}

To demonstrate how the alert tree structure and threat score heuristic may be useful in practice, we include example forward and backward trees in Figures \ref{fig:fwd-tree} and \ref{fig:bwd-tree}, respectively. These trees were selected for their size; some trees produced had well over 1000 vertices and would not be legible in the present format. This problem is a challenge that we leave to future work, as the present focus is efficiency of reconstruction.

\noindent{\bf Answering RQ1}.
In order to answer RQ1, we must retrieve a set of paths corresponding to a known attack origin and a known attack target. Assuming the database maintenance has kept up with the alert stream, this can be accomplished with a query to the database utilizing the $(source,destination)$ index. Because subpaths are also stored in the database, we can be confident that we need only retrieve paths that start and end with the origin and target, respectively. This efficiently returns a list containing exactly the required corresponding paths, without the need to parse the paths to truncate them at the proper destination. 

\noindent{\bf Answering RQ2}.
To answer RQ2, we must reconstruct the forward alert tree corresponding to a particular origin. This process begins with retrieving all of the paths beginning at the specified node, leveraging the $source$ index. We then pass these paths to the tree reconstruction function, which arranges them based on their relationships to each other. 

\noindent{\bf Answering RQ3}.
In order to answer RQ3, we must reconstruct the backward alert tree corresponding to a particular target. Similar to RQ2, this process retrieves the appropriate paths from the database and passes them to the tree reconstruction function. 

\section{Limitations}
\label{sec:discussion}

We identify six limitations that should be addressed in future studies.
First, AutoCRAT depends on IDS correctness, 
meaning that IDS false positives and false negatives can result in errors in AutoCRAT's path and tree reconstruction. However, in the case where errors occur along paths with true positives, the penalty will only reduce the accuracy of the threat score calculations. It is important to build metrics to quantify the impact of IDS (in)correctness on the trustworthiness of the results, in a fashion similar to \cite{XuHotSoS2018Firewall}. In particular, it would be exciting to establish a systematic quantitative methodology that can be seamlessly incorporated into the Cybersecurity Dynamics framework \cite{XuBookChapterCD2019,XuMTD2020} to enable not only reactive defenses but more importantly proactive and adaptive defenses \cite{XuTDSC2011,XuTAAS2012,XuTDSC2012,XuInternetMath2012,XuTAAS2014,XuHotSoS2014-MTD,XuHotSoS2015,XuIEEETNSE2018,XuIEEEACMToN2019,XuTNSE2021-GlobalAttractivity}, by possibly leveraging data-driven cyber threats forecasting techniques \cite{XuSciSec2023-forecasting,XuTIFSSparsity2021,XuJAS2018,XuTIFSDataBreach2018,XuJAS2016,XuTechnometrics2017,XuIEEETIFS2015,XuIEEETIFS2013}.

Second, AutoCRAT 
assumes that attacks are always initiated by a malicious node.
    This assumption may be violated in client-side attacks, such as drive-by downloads \cite{MirIEEECNS2022,XuCNS2014,XuCodaspy13-maliciousURL}. In such cases, an advanced security device may be able to reverse the order of nodes in the alert during preprocessing, preserving attack semantics. It is important to investigate how to extend AutoCRAT to accommodate cyber social engineering attacks and defenses as they are often used as a means to penetrate into a network, especially in relation to the psychological aspect  \cite{XuSciSec2024PF-Evolution,XuSciSec2024PTac-PTech-Evolution,XuPIEEE2024,RosaSocialEngineeringKillChainSciSec2022}. It is also interesting to extend AutoCRAT to smart homes, which is an emerging field especially from the cyber insurance perspective \cite{XuSciSec2024-CyberInsurance}.
    

Third, AutoCRAT depends on the accuracy of threat scores.
This limitation may result in rankings that do not reliably show the importance of a path or tree relative to the mission at hand. However, it does not affect the accuracy of path or tree reconstruction. Regardless, AutoCRAT can be easily adapted to incorporate better ranking methods, which may include asset values and risk scores.
The problem of ranking alerts also remains an open research problem independent of alert path and tree modeling. Thus, we need to develop a systematic family of metrics \cite{Pendleton16,Cho16-milcom,XuSTRAM2018ACMCSUR,XuIEEETIFS2018-groundtruth,XuAgility2019,XuSciSec2021SARR}.
    

Fourth, AutoCRAT's path maintenance algorithm assumes that events are inserted sequentially. This means that it may be difficult to parallelize its execution. Since parallelization is a powerful tool of efficiency, this problem may impact viability of the methods in practice. However, it may be possible to parallelize some alert insertions if the adjacent nodes are disparate relative to sequential alerts. We leave this investigation to future work. 

Fifth,
AutoCRAT assumes that each computer has a single, unique, and static address. It may be extensible to accommodate computers with multiple IP addresses (e.g., by aliasing node names before inserting them into the database or when preprocessing alerts). In the case of a segmented network with private subnets, some computers on disparate subnets may have matching addresses (e.g., 192.168.1.1). This case may be harder to accommodate.

Sixth,
the efficiency of the path maintenance algorithm depends on the assumption that the database is capable of efficiently indexing elements of an array. This restricts the interoperability of the framework to certain kinds of databases.


\section{Conclusion}
\label{sec:conclusion}

We have introduced the AutoCRAT system for modeling and tracking multi-step network attacks as indicated by alerts generated by security devices. The key concepts behind AutoCRAT are those of alert graphs, alert paths, and alert trees. 
The technical contributions include data structures and algorithms for efficiently representing and constructing alert paths and alert trees, as well as asymptotic storage and runtime complexity analysis. This study is useful to cyber defenders because it quantifies threats against a network and its components and presents them in an intuitive form that is easy to understand. Our case study based on an implementation of AutoCRAT and a research dataset shows that AutoCRAT can efficiently reproduce alert paths and trees, keeping pace with alerts produced on a testbed network. 
The paper is a significant step towards automating cyber triage with and risk quantification, which remains an important and elusive problem. More research needs to be conducted with real-world datasets, including cyber-physical systems such as smart homes and hospitals. 

\smallskip

\noindent{\bf Acknowledgement}. We thank the reviewers for their comments. This research was supported in part by NSF Grant \#2115134 and Colorado State Bill 18-086. This research used the Chameleon testbed.

\bibliographystyle{splncs04} 
\bibliography{eric,metrics}

\begin{thebibliography}{10}
\providecommand{\url}[1]{\texttt{#1}}
\providecommand{\urlprefix}{URL }
\providecommand{\doi}[1]{https://doi.org/#1}

\bibitem{suricata}
Suricata | open source ids / ips / nsm engine. \url{https://suricata-ids.org/download/} (2018)

\bibitem{emerging_threats_2019}
Welcome to the emerging threats rule server. \url{https://rules.emergingthreats.net/} (2019)

\bibitem{alsubhi2012fuzmet}
Alsubhi, K., Aib, I., Boutaba, R.: Fuzmet: A fuzzy-logic based alert prioritization engine for intrusion detection systems. International Journal of Network Management  \textbf{22}(4),  263--284 (2012)

\bibitem{apruzzese2017detection}
Apruzzese, G., Pierazzi, F., Colajanni, M., Marchetti, M.: Detection and threat prioritization of pivoting attacks in large networks. IEEE Transactions on Emerging Topics in Computing  \textbf{8}(2),  404--415 (2017)

\bibitem{axelsson2000base}
Axelsson, S.: The base-rate fallacy and the difficulty of intrusion detection. ACM Transactions on Information and System Security (TISSEC)  \textbf{3}(3),  186--205 (2000)

\bibitem{chen2012attack}
Chen, C.M., Guan, D., Huang, Y.Z., Ou, Y.H.: Attack sequence detection in cloud using hidden markov model. In: 2012 seventh asia joint conference on information security. pp. 100--103. IEEE (2012)

\bibitem{XuHotSoS2018Firewall}
Chen, H., Cho, J., Xu, S.: Quantifying the security effectiveness of firewalls and dmzs. In: Proc. HoTSoS'2018. pp. 9:1--9:11 (2018)

\bibitem{chen2007value}
Chen, Y., Boehm, B., Sheppard, L.: Value driven security threat modeling based on attack path analysis. In: 2007 40th Annual Hawaii International Conference on System Sciences (HICSS'07). pp. 280a--280a. IEEE (2007)

\bibitem{Cho16-milcom}
Cho, J., Hurley, P., Xu, S.: Metrics and measurement of trustworthy systems. In: Proc. IEEE MILCOM (2016)

\bibitem{XuSTRAM2018ACMCSUR}
Cho, J.H., Xu, S., Hurley, P.M., Mackay, M., Benjamin, T., Beaumont, M.: Stram: Measuring the trustworthiness of computer-based systems. ACM Comput. Surv.  \textbf{51}(6),  128:1--128:47 (2019)

\bibitem{chou2020data}
Chou, D., Jiang, M.: Data-driven network intrusion detection: A taxonomy of challenges and methods. arXiv preprint arXiv:2009.07352  (2020)

\bibitem{cinque2020contextual}
Cinque, M., Della~Corte, R., Pecchia, A.: Contextual filtering and prioritization of computer application logs for security situational awareness. Future Generation Computer Systems  \textbf{111},  668--680 (2020)

\bibitem{de2018process}
De~Alvarenga, S.C., Barbon~Jr, S., Miani, R.S., Cukier, M., Zarpel{\~a}o, B.B.: Process mining and hierarchical clustering to help intrusion alert visualization. Computers \& Security  \textbf{73},  474--491 (2018)

\bibitem{XuIEEETIFS2018-groundtruth}
Du, P., Sun, Z., Chen, H., Cho, J.H., Xu, S.: Statistical estimation of malware detection metrics in the absence of ground truth. IEEE T-IFS  \textbf{13}(12),  2965--2980 (2018)

\bibitem{XuTIFSSparsity2021}
Fang, Z., Xu, M., Xu, S., Hu, T.: A framework for predicting data breach risk: Leveraging dependence to cope with sparsity. IEEE T-IFS  \textbf{16},  2186--2201 (2021)

\bibitem{GabeIEEEMilcom2019}
Fernandez, G.C., Xu, S.: A case study on using deep learning for network intrusion detection. In: 2019 {IEEE} Military Communications Conference (MILCOM'2019). pp.~1--6 (2018)

\bibitem{XuNSS2022-AlertTree}
Ficke, E., Bateman, R.M., Xu, S.: Reducing intrusion alert trees to aid visualization. In: Yuan, X., Bai, G., Alcaraz, C., Majumdar, S. (eds.) Network and System Security - 16th International Conference, {NSS} 2022, Denarau Island, Fiji, December 9-12, 2022, Proceedings. Lecture Notes in Computer Science, vol. 13787, pp. 140--154. Springer (2022)

\bibitem{ficke2019analyzing}
Ficke, E., Schweitzer, K.M., Bateman, R.M., Xu, S.: Analyzing root causes of intrusion detection false-negatives: Methodology and case study. In: Proc. IEEE MILCOM'2019 (2019)

\bibitem{apin}
Ficke, E., Xu, S.: Apin: Automatic attack path identification in computer networks. In: IEEE ISI 2020 (2020)

\bibitem{mandiant2015numbers}
FireEye: The numbers game: How many alerts is too many to handle? \url https://www.fireeye.com/offers/rpt-idc-numbers-game-special-report.html (2015)

\bibitem{Frigault2008}
{Frigault}, M., {Wang}, L.: Measuring network security using bayesian network-based attack graphs. In: 2008 32nd Annual IEEE International Computer Software and Applications Conference. pp. 698--703 (July 2008). \doi{10.1109/COMPSAC.2008.88}

\bibitem{haas2018gac}
Haas, S., Fischer, M.: Gac: graph-based alert correlation for the detection of distributed multi-step attacks. In: Proceedings of the 33rd Annual ACM Symposium on Applied Computing. pp. 979--988 (2018)

\bibitem{XuHotSoS2014-MTD}
Han, Y., Lu, W., Xu, S.: Characterizing the power of moving target defense via cyber epidemic dynamics. In: HotSoS. pp. 1--12 (2014)

\bibitem{XuTNSE2021-GlobalAttractivity}
Han, Y., Lu, W., Xu, S.: Preventive and reactive cyber defense dynamics with ergodic time-dependent parameters is globally attractive. IEEE TNSE  \textbf{8}(3),  2517--2532 (2021)

\bibitem{hossain2017sleuth}
Hossain, M.N., Milajerdi, S.M., Wang, J., Eshete, B., Gjomemo, R., Sekar, R., Stoller, S., Venkatakrishnan, V.: $\{$SLEUTH$\}$: Real-time attack scenario reconstruction from $\{$COTS$\}$ audit data. In: 26th $\{$USENIX$\}$ Security Symposium ($\{$USENIX$\}$ Security 17). pp. 487--504 (2017)

\bibitem{howard2005measuring}
Howard, M., Pincus, J., Wing, J.M.: Measuring relative attack surfaces. In: Computer security in the 21st century, pp. 109--137. Springer (2005)

\bibitem{hu2018security}
Hu, H., Zhang, H., Yang, Y.: Security risk situation quantification method based on threat prediction for multimedia communication network. Multimedia Tools and Applications  \textbf{77}(16),  21693--21723 (2018)

\bibitem{CyberKillChainPaper2011}
Hutchins, E.M., Cloppert, M.J., Amin, R.M.: Intelligence-driven computer network defense informed by analysis of adversary campaigns and intrusion kill chains. In: 2011 International Conference on Information Warfare and Security (2011)

\bibitem{khraisat2019survey}
Khraisat, A., Gondal, I., Vamplew, P., Kamruzzaman, J.: Survey of intrusion detection systems: techniques, datasets and challenges. Cybersecurity  \textbf{2}(1),  1--22 (2019)

\bibitem{lee2018game}
Lee, S., Kim, S., Choi, K., Shon, T.: Game theory-based security vulnerability quantification for social internet of things. Future Generation Computer Systems  \textbf{82},  752--760 (2018)

\bibitem{leitold2016quantifying}
Leitold, F., Arrott, A., Hadarics, K.: Quantifying cyber-threat vulnerability by combining threat intelligence, it infrastructure weakness, and user susceptibility. In: 24th Annual EICAR Conference, Nuremberg, Germany (2016)

\bibitem{XuTDSC2011}
Li, X., Parker, P., Xu, S.: A stochastic model for quantitative security analyses of networked systems. IEEE TDSC  \textbf{8}(1),  28--43 (2011)

\bibitem{XuIEEEACMToN2019}
Lin, Z., Lu, W., Xu, S.: Unified preventive and reactive cyber defense dynamics is still globally convergent. IEEE/ACM ToN  \textbf{27}(3),  1098--1111 (2019)

\bibitem{XuPIEEE2024}
Longtchi, T., Rodriguez, R.M., Al{-}Shawaf, L., Atyabi, A., Xu, S.: Internet-based social engineering psychology, attacks, and defenses: A survey. Proceedings of IEEE  \textbf{112}(3),  210--246 (2024)

\bibitem{XuSciSec2024PF-Evolution}
Longtchi, T., Xu, S.: Characterizing the evolution of psychological factors exploited by malicious emails. In: Proceedings of International Conference on Science of Cyber Security (SciSec'2024) (2024)

\bibitem{XuSciSec2024PTac-PTech-Evolution}
Longtchi, T., Xu, S.: Characterizing the evolution of psychological tactics and techniques exploited by malicious emails. In: Proceedings of International Conference on Science of Cyber Security (SciSec'2024) (2024)

\bibitem{Mandiant}
Mandiant: Apt1 report. \url{https://www.fireeye.com/content/dam/fireeyewww/services/pdfs/mandiant-apt1-report.pdf} (2013)

\bibitem{mao2021mif}
Mao, B., Liu, J., Lai, Y., Sun, M.: Mif: A multi-step attack scenario reconstruction and attack chains extraction method based on multi-information fusion. Computer Networks  \textbf{198},  108340 (2021)

\bibitem{XuAgility2019}
Mireles, J., Ficke, E., Cho, J., Hurley, P., Xu, S.: Metrics towards measuring cyber agility. IEEE Transactions on Information Forensics and Security  \textbf{14}(12),  3217--3232 (2019)

\bibitem{ning2002constructing}
Ning, P., Cui, Y., Reeves, D.S.: Constructing attack scenarios through correlation of intrusion alerts. In: Proceedings of the 9th ACM Conference on Computer and Communications Security. pp. 245--254 (2002)

\bibitem{ning2003learning}
Ning, P., Xu, D.: Learning attack strategies from intrusion alerts. In: Proceedings of the 10th ACM conference on Computer and communications security. pp. 200--209 (2003)

\bibitem{ou2006scalable}
Ou, X., Boyer, W.F., McQueen, M.A.: A scalable approach to attack graph generation. In: Proceedings of the 13th ACM conference on Computer and communications security. pp. 336--345 (2006)

\bibitem{Pendleton16}
Pendleton, M., Garcia-Lebron, R., Cho, J.H., Xu, S.: A survey on systems security metrics. ACM Comput. Surv.  \textbf{49}(4),  62:1--62:35 (Dec 2016)

\bibitem{XuJAS2016}
Peng, C., Xu, M., Xu, S., Hu, T.: Modeling and predicting extreme cyber attack rates via marked point processes. Journal of Applied Statistics  \textbf{44}(14),  2534--2563 (2017)

\bibitem{XuJAS2018}
Peng, C., Xu, M., Xu, S., Hu, T.: Modeling multivariate cybersecurity risks. Journal of Applied Statistics  \textbf{0}(0),  1--23 (2018)

\bibitem{phillips1998graph}
Phillips, C., Swiler, L.P.: A graph-based system for network-vulnerability analysis. In: Proceedings of the 1998 workshop on New security paradigms. pp. 71--79 (1998)

\bibitem{MirIEEECNS2022}
Pritom, M., Xu, S.: Supporting law-enforcement to cope with blacklisted websites: Framework and case study. In: IEEE CNS'2022 (2022)

\bibitem{ramaki2018systematic}
Ramaki, A.A., Rasoolzadegan, A., Bafghi, A.G.: A systematic mapping study on intrusion alert analysis in intrusion detection systems. ACM Computing Surveys (CSUR)  \textbf{51}(3),  1--41 (2018)

\bibitem{RosaSocialEngineeringKillChainSciSec2022}
Rodriguez, R.M., Xu, S.: Cyber social engineering kill chain. In: Proceedings of International Conference on Science of Cyber Security (SciSec'2022). pp. 487--504 (2022)

\bibitem{ibm2021breachreport}
Security, I.: Cost of a data breach report 2021. \url{https://www.ibm.com/downloads/cas/OJDVQGRY} (July 2021)

\bibitem{sharafaldin2018toward}
Sharafaldin, I., Lashkari, A.H., Ghorbani, A.A.: Toward generating a new intrusion detection dataset and intrusion traffic characterization. In: ICISSP. pp. 108--116 (2018)

\bibitem{ATTCK}
Strom, B.: Att\&ck 101: Cyber threat intelligence (2018), \url{{https://www.mitre.org/capabilities/cybersecurity/overview/cybersecurity-blog/attck-101}}

\bibitem{XuSciSec2023-forecasting}
Sun, Z., Xu, M., Schweitzer, K., Bateman, R., Kott, A., Xu, S.: Cyber attacks against enterprise networks: Characterization, modeling and forecasting. In: Proc. of SciSec'2023 (2023)

\bibitem{thakkar2020review}
Thakkar, A., Lohiya, R.: A review of the advancement in intrusion detection datasets. Procedia Computer Science  \textbf{167},  636--645 (2020)

\bibitem{vasilomanolakis2016towards}
Vasilomanolakis, E., Cordero, C.G., Milanov, N., M{\"u}hlh{\"a}user, M.: Towards the creation of synthetic, yet realistic, intrusion detection datasets. In: NOMS 2016-2016 IEEE/IFIP Network Operations and Management Symposium. pp. 1209--1214. IEEE (2016)

\bibitem{wang2021maac}
Wang, X., Gong, X., Yu, L., Liu, J.: Maac: Novel alert correlation method to detect multi-step attack. In: 2021 IEEE 20th International Conference on Trust, Security and Privacy in Computing and Communications (TrustCom). pp. 726--733. IEEE (2021)

\bibitem{XuCNS2014}
Xu, L., Zhan, Z., Xu, S., Ye, K.: An evasion and counter-evasion study in malicious websites detection. In: IEEE CNS. pp. 265--273 (2014)

\bibitem{XuCodaspy13-maliciousURL}
Xu, L., Zhan, Z., Xu, S., Ye, K.: Cross-layer detection of malicious websites. In: ACM CODASPY'13. pp. 141--152 (2013)

\bibitem{XuTechnometrics2017}
Xu, M., Hua, L., Xu, S.: A vine copula model for predicting the effectiveness of cyber defense early-warning. Technometrics  \textbf{59}(4),  508--520 (2017)

\bibitem{XuTIFSDataBreach2018}
Xu, M., Schweitzer, K.M., Bateman, R.M., Xu, S.: Modeling and predicting cyber hacking breaches. IEEE T-IFS  \textbf{13}(11),  2856--2871 (2018)

\bibitem{XuInternetMath2012}
Xu, M., Xu, S.: An extended stochastic model for quantitative security analysis of networked systems. Internet Mathematics  \textbf{8}(3),  288--320 (2012)

\bibitem{XuMTD2020}
Xu, S.: The cybersecurity dynamics way of thinking and landscape (invited paper). In: ACM Workshop on Moving Target Defense (2020)

\bibitem{XuTAAS2012}
Xu, S., Lu, W., Xu, L.: Push- and pull-based epidemic spreading in networks: Thresholds and deeper insights. ACM TAAS  \textbf{7}(3) (2012)

\bibitem{XuTAAS2014}
Xu, S., Lu, W., Xu, L., Zhan, Z.: Adaptive epidemic dynamics in networks: Thresholds and control. ACM TAAS  \textbf{8}(4) (2014)

\bibitem{XuTDSC2012}
Xu, S., Lu, W., Zhan, Z.: A stochastic model of multivirus dynamics. IEEE Transactions on Dependable and Secure Computing  \textbf{9}(1),  30--45 (2012)

\bibitem{XuBookChapterCD2019}
Xu, S.: Cybersecurity dynamics: A foundation for the science of cybersecurity. In: Lu, Z., Wang, C. (eds.) Proactive and Dynamic Network Defense, vol.~74, pp. 1--31. Springer Nature Switzerland AG (2019)

\bibitem{XuSciSec2021SARR}
Xu, S.: Sarr: A cybersecurity metrics and quantification framework. In: Third International Conference on Science of Cyber Security (SciSec'2021). pp. 3--17 (2021)

\bibitem{XuIEEETIFS2013}
Zhan, Z., Xu, M., Xu, S.: Characterizing honeypot-captured cyber attacks: Statistical framework and case study. IEEE Transactions on Information Forensics and Security  \textbf{8}(11),  1775--1789 (2013)

\bibitem{XuIEEETIFS2015}
Zhan, Z., Xu, M., Xu, S.: Predicting cyber attack rates with extreme values. {IEEE} Transactions on Information Forensics and Security  \textbf{10}(8),  1666--1677 (2015)

\bibitem{zhang2019intrusion}
Zhang, K., Zhao, F., Luo, S., Xin, Y., Zhu, H.: An intrusion action-based ids alert correlation analysis and prediction framework. IEEE Access  \textbf{7},  150540--150551 (2019)

\bibitem{XuSciSec2024-CyberInsurance}
Zhang, X., Xu, M., Xu, S.: Smart home cyber insurance pricing. In: Proceedings of International Conference on Science of Cyber Security (SciSec'2024) (2024)

\bibitem{XuHotSoS2015}
Zheng, R., Lu, W., Xu, S.: Active cyber defense dynamics exhibiting rich phenomena. In: Proc. HotSoS (2015)

\bibitem{XuIEEETNSE2018}
Zheng, R., Lu, W., Xu, S.: Preventive and reactive cyber defense dynamics is globally stable. IEEE TNSE  \textbf{5}(2),  156--170 (2018)

\end{thebibliography}

\end{document}